\documentclass{llncs}
\usepackage[T1]{fontenc}
\usepackage{graphicx}
\usepackage{textcomp}
\usepackage{array,ragged2e}

\def\BibTeX{{\rm B\kern-.05em{\sc i\kern-.025em b}\kern-.08em
    T\kern-.1667em\lower.7ex\hbox{E}\kern-.125emX}}

\usepackage{colortbl}
\usepackage{makecell}
\usepackage{pgfplots}
\pgfplotsset{compat=1.17}
\usepackage{multicol}
\usepackage{multirow}
\usepackage{tabto}
\usepackage{wrapfig}
\usepackage{amssymb}
\usepackage{float}
\usepackage[misc,geometry]{ifsym}
\usepackage{tabularx}
\usepackage{supertabular}
\usepackage{longtable}
\usepackage{graphicx}
\usepackage{subcaption}
\usepackage{textcomp}
\usepackage{array,ragged2e}
\usepackage{pgfplots}
\usepackage{url}
\usepackage{lineno,hyperref}
\hypersetup{
  linkcolor  = red!60!black,
  citecolor  = blue!90!white,
  urlcolor   = violet!60!black,
  colorlinks = true,
}
\usepackage{orcidlink}
\usepackage{hyperref}
\usepackage{svg}
\usepackage{xcolor}
\usepackage{threeparttable}

\begin{document}
\title{A Graph Based Raman Spectral Processing Technique for Exosome Classification}

\author{
Vuong M. Ngo\orcidlink{0000-0002-8793-0504}\inst{1,4}~\textsuperscript{\Letter}
\and
Edward Bolger\inst{2}
\and
Stan Goodwin\inst{2}
\and\\
John O’Sullivan\inst{3}
\and
Dinh Viet Cuong \inst{1}
\and
Mark Roantree\orcidlink{0000-0002-1329-2570}\inst{1}
}
\authorrunning{Ngo, V.M. et al.}

\institute{
Insight Centre for Data Analytics, Dublin City University, Ireland  \and
School of Computing, Dublin City University, Ireland \and
School of Chemical Sciences, Dublin City University, Ireland \and
Ho Chi Minh City Open University, Ho Chi Minh City, Vietnam
\\
\email{vuong.ngo@dcu.ie or vuong.nm@ou.edu.vn, 
    eddie.bolger62@gmail.com, stangoodwin02@gmail.com, 
    johns@gmail.com,
    mark.roantree@dcu.ie}
}

\maketitle          
\begin{abstract}
Exosomes are small vesicles crucial for cell signaling and disease biomarkers. Due to their complexity, an "omics" approach is preferable to individual biomarkers. While Raman spectroscopy is effective for exosome analysis, it requires high sample concentrations and has limited sensitivity to lipids and proteins. Surface-enhanced Raman spectroscopy helps overcome these challenges. In this study, we leverage Neo4j graph databases to organize 3,045 Raman spectra of exosomes, enhancing data generalization. To further refine spectral analysis, we introduce a novel spectral filtering process that integrates the PageRank Filter with optimal Dimensionality Reduction. This method improves feature selection, resulting in superior classification performance. Specifically, the Extra Trees model, using our spectral processing approach, achieves 0.76 and 0.857 accuracy in classifying hyperglycemic, hypoglycemic, and normal exosome samples based on Raman spectra and surface, respectively, with group 10-fold cross-validation. Our results show that graph-based spectral filtering combined with optimal dimensionality reduction significantly improves classification accuracy by reducing noise while preserving key biomarker signals. This novel framework enhances Raman-based exosome analysis, expanding its potential for biomedical applications, disease diagnostics, and biomarker discovery.

\end{abstract}

% \textsuperscript{$\ddagger$}{These authors are equal contributors to this work.}
%
%
%
\section{Introduction}

Traditional diagnosis of cellular diseases such as cancer, diabetes, hyperglycemia and hypoglycemia can be a complex and invasive task. Cancer diagnosis is typically carried out using tissue biopsies, where tissue is extracted from a suspected tumour and analysed in a lab \cite{sierra2020}. This process is time consuming and expensive and can potentially pose risks to the patient. Likewise, current approaches for early diabetes screening are reliant on blood tests, which reduces the likelihood of patients receiving optimal treatment plans \cite{sun2021}. As a result, there is a significant drive to discover faster, less invasive methods of disease screening.

\begin{wrapfigure}{r}{0.45\textwidth}
    \centering
    \vspace{-3mm}
    \includegraphics[width=1.0\linewidth]{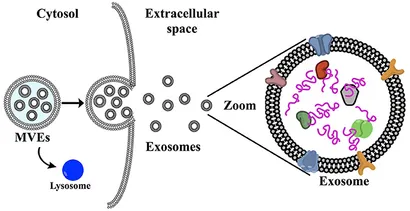}
    \caption{The Emission of Exosomes from a Cell \cite{delatorre2018}}
    \label{fig:exosomes}
\end{wrapfigure}

Exosomes are microscopic extracellular vesicles that play a crucial role in cell-to-cell communication, as illustrated in Figure \ref{fig:exosomes}. They carry proteins, DNA, RNA, and other biological materials derived from the cells that produce them \cite{kalluri2020}. Exosomes are present in nearly all bodily fluids, including blood, saliva, and urine \cite{qin2014}. Their composition makes them valuable as biomarkers, revealing the presence of diseases in their parent cells \cite{sun2021}. Because exosomes can be obtained from common bodily fluids, non-invasive liquid biopsies offer significant advantages over traditional tissue biopsies, such as reduced patient stress and quicker diagnostic processes. These liquid biopsies have been successfully applied to diagnose various diseases, including prostate, pancreatic, breast, and ovarian cancers, as well as diabetes and hyperglycemia \cite{kalluri2016}, \cite{kopec2022}.

Due to their small size (40–150 nm) and presence in complex biological samples, exosomes require specialized techniques for measurement and analysis \cite{li2022}. Raman Spectroscopy, a non-destructive vibrational spectroscopic technique based on inelastic photon scattering, is particularly effective for analyzing exosome structures \cite{das2011}. Using a monochromatic laser, it excites chemical bonds within the particle, causing them to emit radiation with a slight frequency shift, known as the Raman shift, which forms the x-axis of a Raman spectrum. The spectrum’s peaks represent specific molecular bond vibrations, creating a unique "fingerprint" that identifies a substance's shape and composition. Surface-Enhanced Raman Spectroscopy (SERS) further amplifies the Raman signal by placing the sample on roughened metal surfaces, enhancing its detection sensitivity.

%----------------------

%% REMOVED
%As spectral peaks correspond to specific exosomal biomarkers \cite{osullivan2022}, our initial approach utilized peak profiling and graph databases to extract features that more effectively capture the distinguishing characteristics of each spectrum. 

\vspace{2mm}
\textbf{Contribution.}
The contribution in this paper can be articulated as follows:
\vspace{-3mm}
\begin {itemize}
    \item To enhance the identification of representative spectra, we developed a novel spectral filtering process that integrates the PageRank Filter (RDF) with optimal Dimensionality Reduction (DR) centrality. This method removes noisy or unrepresentative spectra while preserving the most informative spectral features, leading to a more robust dataset for classification.
    \item We applied a range of machine learning models to classify exosome samples based on processed Raman spectra or surfaces. The models were chosen for their strong performance in high-dimensional data and their ability to capture complex relationships within spectral features.
    \item To ensure the generalizability of our approach, we employed group $k$-fold cross-validation, which prevents data leakage by ensuring that spectra from the same experimental surface do not appear in both training and test sets.
\end {itemize}

\textbf{Paper Structure.}
The remainder of this paper is organized as follows: Section \ref{sec:related_work} reviews related work. Section \ref{sec:methodology} introduces our novel Raman spectra processing method, highlighting key challenges in Raman spectra and how our approach addresses them. Section \ref{sec:experiment} presents the experimental results and analysis of the proposed spectral processing techniques. Finally, Section \ref{sec:conclusion} concludes the paper and outlines potential directions for future research.

\section{Related Work}
\label{sec:related_work}
Most studies applying machine learning to Raman spectra follow a standard preprocessing sequence: despiking to remove cosmic ray artifacts, baseline correction to isolate the Raman signal, smoothing to reduce noise, and normalization to adjust for intensity variations \cite{hu2022}, \cite{chen2023}, \cite{zhang2022}. To mitigate dimensionality where large feature spaces can degrade classifier performance, feature reduction techniques such as principal component analysis (PCA), linear discriminant analysis (LDA), or partial least squares (PLS) are applied \cite{schumacher2014}. Machine learning is typically performed using algorithms like Support Vector Machines (SVM) \cite{zhang2022}, \cite{amjad2018}.

Shin et al. \cite{shin2020} employed a Residual Neural Network to classify Raman spectra of exosomes derived from normal and lung cancer cell lines. Their study utilized standard preprocessing steps, including baseline correction, denoising, and normalization. Neural networks, unlike traditional models, can automatically extract features from the data, eliminating the need for manual feature engineering. Similarly, Xie et al. \cite{xie2022} used an Artificial Neural Network (ANN) with SERS spectra of exosomes to successfully detect four types of breast cancer.

Utilising Graph Representations with Machine Learning, Graph databases have been successfully used to develop additional machine learning features in relation to disease classification. In a paper by Alqaissi et al. \cite{alqaissi2023}, they were able to use a knowledge graph constructed on COVID-19 literature to train a random forest model to detect COVID-19 based on symptoms. This graph-based random forest model outperformed other models that used the same dataset. Furthermore, they found that incorporating Fast Random Projection (FastRP) node embeddings as features, further improved the performance of the model. According to the researcher, the study “demonstrates that graph algorithms support extracting essential features from the COVID-19 dataset”. 

Graph representations have been effectively utilized to enhance machine learning features for disease classification and spectral analysis. Alqaissi et al. \cite{alqaissi2023} used a knowledge graph derived from COVID-19 literature to train a random forest model for detecting COVID-19 based on symptoms,incorporating Fast Random Projection (FastRP) node embeddings to improve performance. Similarly, Wang et al. \cite{wang2021} applied graph representations to classify Raman spectra of oil paper. They constructed graphs by representing spectra as 1023-dimensional vectors based on consistent wave number intervals, calculating Euclidean distances, and applying a Gaussian kernel to measure similarity. This enabled the creation of a Graph Convolutional Network, which successfully classified spectra samples.

%---------------------------------------------------
\section{Methodology}
\label{sec:methodology}
Raman spectra were generated to classify exosomes from normal, hyperglycemic, and hypoglycemic cells using a Horiba LabRAM HR spectrophotometer with a Leica 50X (0.55 N.A.) long-distance objective lens. Spectra were recorded in the 200 to 2000 cm\textsuperscript{-1} range using a 785 nm laser (60.2 mW) with a 10\% O.D. filter. Each acquisition lasted 1 second, with a 4-second exposure time and 9 frame accumulations. The instrument was calibrated according to the manufacturer’s guidelines using the Rayleigh line of a silicon wafer \cite{osullivan2022}. Our dataset includes 3,045 Raman spectra, with fields SpecID, WaveNumber, Absorbance, SurID, and Status. It contains 63 unique SERS surfaces and categorizes the spectra into 915 hyperglycemic, 1,065 hypoglycemic, and 1,065 normal samples.

\subsection{Graph Data Modelling}
Our novel method for reflecting the non-uniform nature of the peaks used graph databases. Centrality algorithms were run on many graphs constructed using Neo4j \cite{neo4j} through Python, using the approach in \cite{10382155} to quantitatively measure the importance of a node in the exosome network. These measures used were: PageRank \cite{brin2012}, Degree \cite{gleich2015}, Eigenvector \cite{zhang2017} and Article Rank centrality \cite{ruhnau2000}. We then ran a node embedding algorithm, FastRP \cite{chen2019}, to capture high-dimensional aspects of the graph structure in a dense, lower dimensional vector \cite{zhou2022}.

Peak Grids were used in designing the graph. Specifically, each peak within a spectral sample is represented as a node. Connections (edges) between nodes are established under two conditions: (1) if two peaks originate from the same spectrum or (2) if two peaks are located close to one another in proximity, suggesting that they might correspond to the same biomarker \cite{das2011}. Our final graph database consists of 27,666 nodes, 4,142,814 edges, and 4,336,476 properties. This approach leverages the spatial relationship and co-occurrence of peaks to identify potential biomarkers, ensuring a more comprehensive representation of the underlying biological patterns within the data. 

%-----------------------------------------------
\vspace{-3mm}
\subsection{OSC: Optimal Spectra Cleaning Method}
Before classification, preprocessing spectral data is crucial to eliminate background noise that may obscure the Raman signal. Our OSC process consists of four main steps: cosmic spike removal, baseline correction, smoothing, and scaling. First, we applied the algorithm by Whitaker and Hayes \cite{whitaker2018} to remove artifacts caused by cosmic ray interference, preserving spectral integrity. Next, baseline correction was performed using the asymmetric least squares algorithm by Eilers and Boelens \cite{eilers2005}, which effectively removes background fluorescence through asymmetric weighting of deviations from the smooth graph. This method is particularly advantageous as it requires no prior knowledge of peak shapes.

To minimize high-frequency noise, the Savitzky-Golay filter, a robust digital signal processing technique widely used in spectral analysis to preserve signal characteristics while reducing noise \cite{gallagher2020}. Finally, scaling was applied to account for variations in absolute Raman intensity, ensuring the focus remains on the intrinsic composition of the exosome rather than its quantity in the sample \cite{osullivan2022}. This step also mitigates variability from external factors, such as fluctuations in laser intensity \cite{martyna2020}. Three normalization methods were evaluated: scaling to the maximum peak, vector norm scaling, and Standard Normal Variate (SNV) scaling.

\begin{figure}[ht]
    \centering
    % First image
    \begin{subfigure}[b]{0.44\textwidth}
        \centering
        \includegraphics[width=\textwidth]{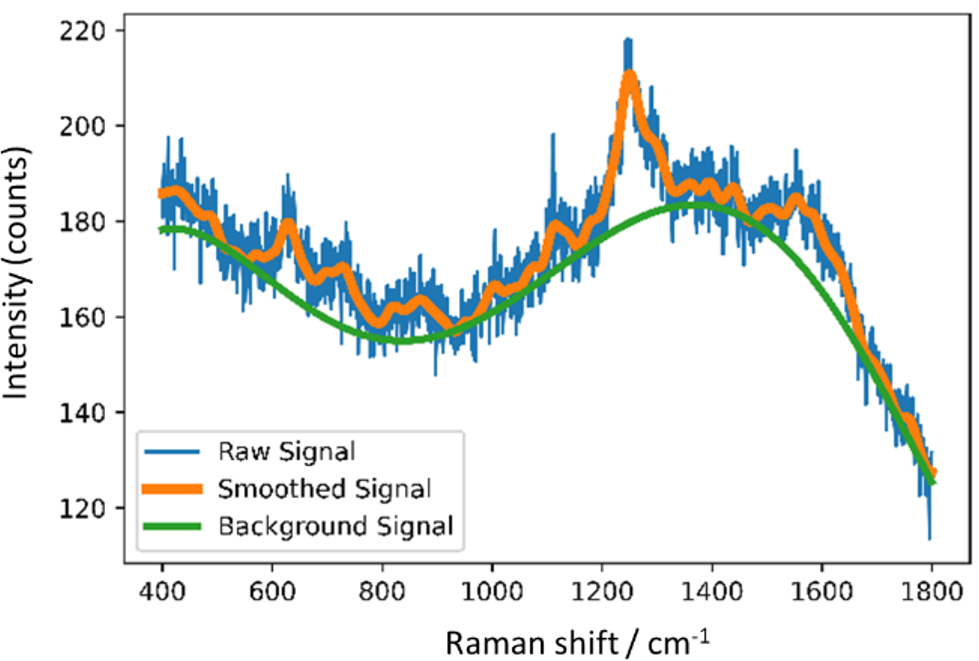}
        \caption{A Raman spectrum at steps}
        \label{fig:smoothed-signal}
    \end{subfigure}
    \hfill
    % Second image
    \begin{subfigure}[b]{0.54\textwidth}
        \centering
        \includegraphics[width=\textwidth]{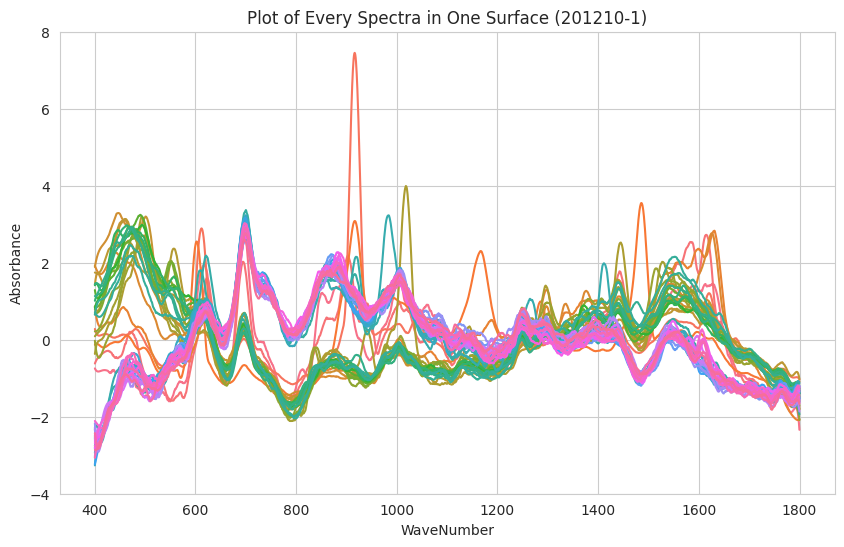}
        \caption{Normalized spectra from the same surface}
        \label{fig:spectra_same_surface}
    \end{subfigure}
    \caption{Examples of Raman spectra after OSC proccesing}
    \label{fig:OSC-Processing}
\end{figure}

Figure \ref{fig:smoothed-signal} presents the raw confocal SERS data (blue) obtained from an exosome-captured gold cavity array, alongside its smoothed version (orange) and baseline-corrected version (green). Meanwhile,  Figure \ref{fig:spectra_same_surface} displays normalized spectra from the same surface, highlighting both representative spectra and a few noisier outliers.

\vspace{-3mm}
\vspace{-5mm}

However, as seen in Figure \ref{fig:spectra_same_surface}, several spectra overlap, suggesting a shared underlying Raman fingerprint. Additionally, some samples deviate from the expected representative spectra, likely due to surface noise or contamination. To assess whether these outliers negatively impacted model performance, we developed methods to identify and remove them.

%------------------------------ 
\subsection{OSC+PRF Method: Enhance OSC using a PageRank Filter}

Spectral readings are typically averaged to reduce background interference \cite{hu2022}, \cite{gautam2015}. However, this step was not performed in our dataset, meaning we retained noisy and unrepresentative samples. To extract only relevant spectra from each surface, we first employed an interquartile range (IQR) method for spectral intensity. An outlier was defined as any value outside 1.5 times the IQR. Spectra were discarded if a certain proportion of their values were classified as outliers. We systematically tested different proportion cutoffs, which led to a slight improvement in performance and motivated us to explore additional outlier removal techniques.

\begin{figure}[H]
    \centering
    \includegraphics[width=1\linewidth]{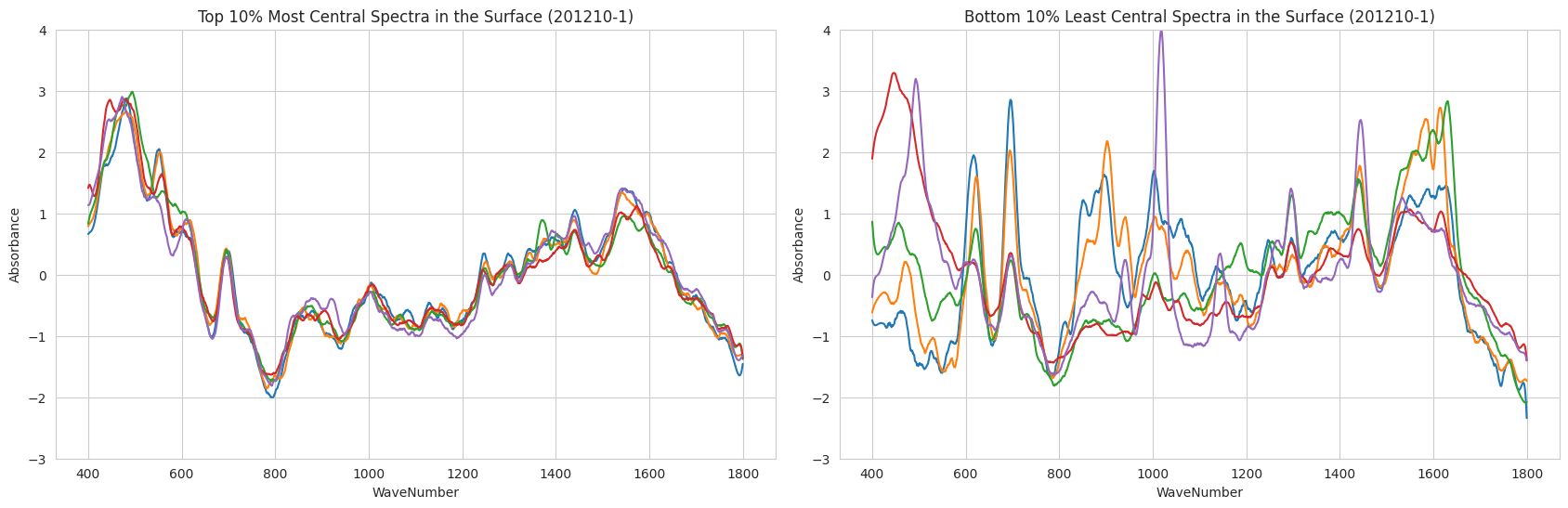}
    \caption{The 10\%  most and least central spectra within a surface according to OSC+PRF processing}
    \label{fig:OSC+PRF}
\end{figure}

Results were then analyzed using a Gaussian Kernel approach. After normalizing the spectra using SNV scaling to ensure that similar shapes were assigned high similarity, we constructed a Gaussian kernel subgraph for each surface. PageRank centrality was then computed for each spectral node, under the assumption that spectra with higher centrality scores better represent the underlying Raman signal, while those with low centrality scores indicate noise or irrelevant data. Figure \ref{fig:OSC+PRF} illustrates this concept, showing the most and least central spectra. We applied a PageRank cutoff to filter out spectra below a certain centrality threshold.

\vspace{-3mm}
\vspace{-5mm}

%-------------------------
\vspace{-3mm}
\subsection{OSC+PRF+DR Method: Applying Optimal Dimensionality Reduction to OSC+PRF}

\vspace{-3mm}
\begin{figure}[H]
    \centering
    \includegraphics[width=0.74\linewidth]{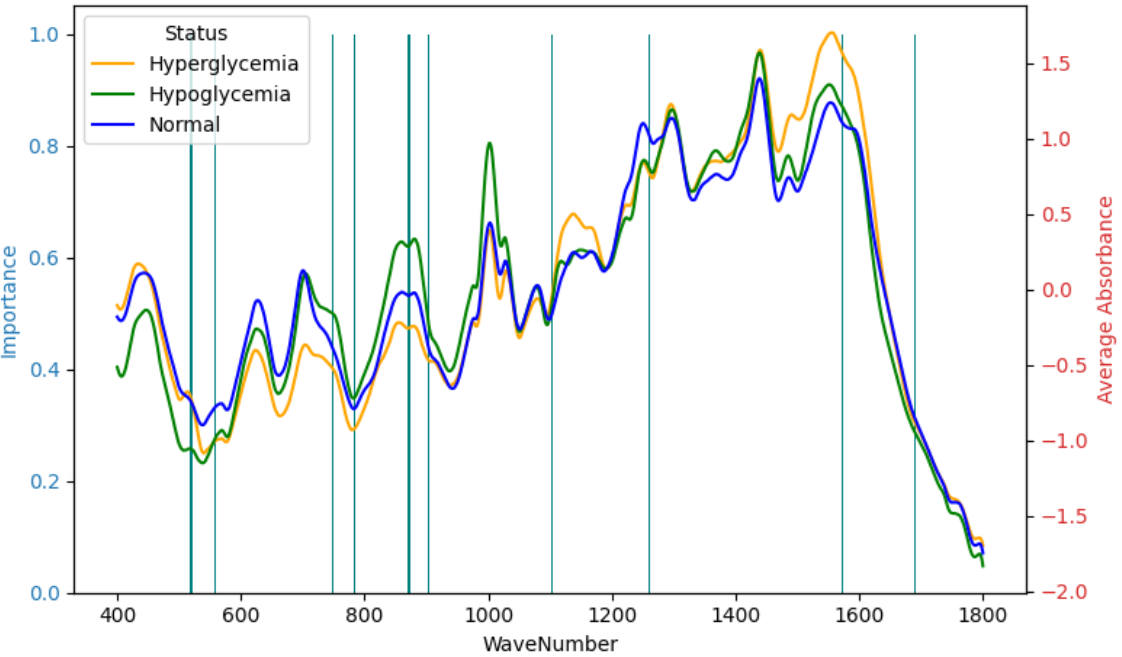}
    \caption{Average spectra by status via OSC+PRF+DR processing}
    \label{fig:OSC+PRF+DR}
\end{figure}
\vspace{-5mm}

To address dimensionality, we explored various feature extraction and selection techniques. Initially, methods such as PCA, LDA, and limiting wave numbers to predefined regions of interest were tested. As each approach resulted in suboptimal model performance, it was necessary to refine the feature set by employing a forward sequential feature selection process, iteratively selecting the wave numbers that contributed most to model performance. Figure \ref{fig:OSC+PRF+DR} presents the average spectra for hyperglycemia, hypoglycemia, and normal conditions after OSC+PRF+DR processing. This approach significantly improved model performance by focusing on the most relevant spectral features. Given the high computational cost of evaluating all possible feature combinations, we limited the selection process to the 50 most influential features, balancing performance gains with computational efficiency.
%--------------------------------
\vspace{-3mm}
\section{Evaluation and Discussion}
\label{sec:experiment}
\vspace{-2mm}
\subsection{Experimental Setup}

To evaluate the performance of our models, we used Group 10-fold cross-valida\-tion, as Guo et al. \cite{guo2017} emphasized the risks of using standard $k$-fold cross-validation with Raman spectra. They recommended splitting data at the highest hierarchical level, such as by biological or technical replicates, to ensure an unbiased estimate of the generalization error. Group $k$-fold cross-validation prevents data leakage by ensuring that samples from the same group are not included in both the training and validation sets, thereby providing a more reliable estimate of the model's generalization performance.

Finally, average accuracy, precision, recall, and F1-score across all folds were calculated to evaluate and compare the overall performance of the models. These metrics provide a comprehensive assessment, with accuracy reflecting the proportion of correct predictions, precision measuring the ability to avoid false positives, recall assessing the ability to identify true positives, and the F1-score offering a balanced evaluation by combining precision and recall.

%--------------------------
\vspace{-3mm}
\subsection{Results}

\vspace{-8mm}
\begin{table}
\centering
\caption{Classification Performance for Raman Spectra Analysis}
\label{tab:spectra_performance}
    \begin{tabular}{l|m{23mm}|r|r|r|r} 
         \textbf{Processing Method} &  Best Model &  Accuracy &  Precision &  Recall & F1-Score \\ \hline
         OSC &  Random Forest &  0.635 &  0.641 &  0.648 &  0.621 \\ 
         OSC+PRF &  Extra Trees &  0.696 &  0.609 &  0.644 &  0.606 \\ 
         OSC+PRF+DR &  Extra Trees &  0.760 &  0.675 &  0.708 &  0.675 \\ 
    \end{tabular}
\end{table}
\vspace{-5mm}

Table \ref{tab:spectra_performance} presents the classification performance of various approaches for Raman spectra analysis using group 10-fold cross-validation. Three methods, OSC, OSC+PRF, and OSC+PRF+DR, were evaluated using different models, namely SVM, Random Forest, and Extra Trees. The results indicate that the Extra Trees classifier performed best for OSC+PRF and OSC+PRF+DR, while Random Forest was the optimal choice for OSC. Among the three approaches, OSC+PRF+DR demonstrated the highest performance, achieving an accuracy of 0.760, precision of 0.675, recall of 0.708, and an F1-score of 0.675, outperforming both OSC+PRF and OSC.

Table \ref{tab:surface_performancel} presents the classification performance results for Raman surface analysis using group 10-fold cross-validation. As with the spectra analysis, best-performing models were Random Forest for OSC and Extra Trees for OSC+PRF and OSC+PRF+DR. A similar trend was observed, with OSC+PRF+DR outperforming OSC+PRF and OSC. However, all approaches demonstrated higher performance compared to spectra analysis. OSC+PRF+DR achieved the highest performance across all metrics, with an accuracy of 0.857, precision of 0.873, recall of 0.864, and an F1-score of 0.857.

\vspace{-5mm}
\begin{table}
\centering
\caption{Classification performance of approaches for Raman surface analysis}
\label{tab:surface_performancel}
    \begin{tabular}{l|m{23mm}|r|r|r|r}
         \textbf{Processing Method} & Best Model &  Accuracy &  Precision &  Recall & F1-Score \\ \hline
         OSC & Random Forest &  0.683 &  0.700 &  0.685 &  0.687 \\ 
         OSC+PRF & Extra Trees &  0.825 &  0.838 &  0.831 &  0.824 \\ 
         OSC+PRF+DR & Extra Trees & 0.857 &  0.873 &  0.864 &  0.857 \\ 
    \end{tabular}
\end{table}
\vspace{-5mm}

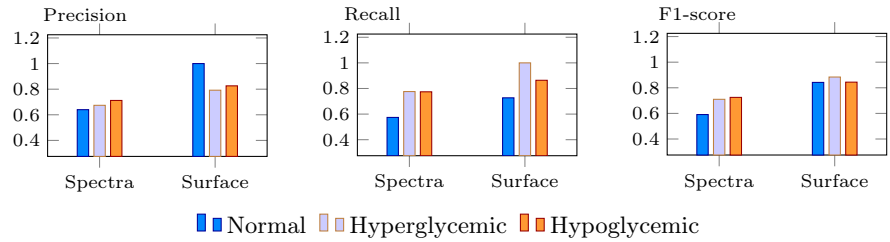
\begin{figure}[H]
\scriptsize
 \begin{center}
  \begin{tikzpicture}
   \pgfplotsset{height=32mm, width= 45mm, xlabel near ticks, ylabel near ticks, 
    plot 1/.style={blue!60!black,fill={rgb,255:red,0; green,135; blue,255},mark=none},%
    plot 2/.style={brown,fill=blue!20!white,mark=none},%
    plot 3/.style={red!60!black,fill={rgb,255:red,255; green,153; blue,51},mark=none}%
   }
    
    %-----------------
    %Precision
   \begin{axis}[
    name=left axis,
    ybar,
    bar width=4.3pt,
    enlargelimits=0.45,
    legend style={/tikz/every even column/.append style={column sep=0.1cm},
    draw=none, fill=none, font=\footnotesize, legend columns=-1, anchor = south, xshift= 25mm, yshift= -28mm}, 
    y label style={align=center},
    ylabel={Precision},
    every axis y label/.style={
    	at={(ticklabel* cs:1.05)},
    	anchor=south, xshift= 5mm,
    },
    ymin=0.5, ymax=1.0,
    symbolic x coords={Spectra, Surface},
    xtick=data,
    % nodes near coords,
    % nodes near coords align={top},
    every node near coord/.append style={font=\scriptsize},  
    ]
    \addplot[plot 1] coordinates{(Spectra,0.64) (Surface,1.0)};
    \addplot[plot 2] coordinates{(Spectra,0.674) (Surface,0.792)};
    \addplot[plot 3] coordinates{(Spectra,0.712) (Surface,0.826)};

	\legend{Normal, Hyperglycemic, Hypoglycemic}
   \end{axis}
   
   %%%%%%%%%%%%%%%%%%%%%%%%%%%%%%%%%%%%%%%%%%%%%%%%%%%%%%%%%
    %Recall
   \begin{axis}[
    name=center axis,
    at=(left axis.east),
    xshift= 1.2cm,
	yshift= -0.8cm,
    ybar,
    bar width=4.3pt,
    enlargelimits=0.45,
    legend style={/tikz/every even column/.append style={column sep=0.25cm},
    draw=none, fill=none, font=\footnotesize, legend columns=-1,
     anchor = north}, 
    y label style={align=center},
    ylabel={Recall},
    every axis y label/.style={
    	at={(ticklabel* cs:1.05)},
    	anchor=south, xshift= 2mm,
    },
    ymin=0.5, ymax=1.0,
    symbolic x coords={Spectra, Surface},
    xtick=data,
    % nodes near coords,
    % nodes near coords align={vertical},
    every node near coord/.append style={font=\scriptsize},	
    ]
    \addplot[plot 1] coordinates{(Spectra,0.574) (Surface,0.727)};
    \addplot[plot 2] coordinates{(Spectra,0.776) (Surface,1.0)};
    \addplot[plot 3] coordinates{(Spectra,0.774) (Surface,0.864)};
   \end{axis}
   
   %------------------------------
   %F1-Score
   \begin{axis}[
    name=right axis,
    at=(center axis.east),
    xshift= 1.2cm,
	yshift= -0.8cm,
    ybar,
    bar width=4.3pt,
    enlargelimits=0.45,
    legend style={/tikz/every even column/.append style={column sep=0.25cm},
    draw=none, fill=none, font=\footnotesize, legend columns=-1,
     anchor = north, xshift= 5mm}, 
    y label style={align=center},
    ylabel={F1-score},
    every axis y label/.style={
    	at={(ticklabel* cs:1.05)},
    	anchor=south, xshift= 4mm,
    },
    ymin=0.5, ymax=1.0,
    symbolic x coords={Spectra, Surface},
    xtick=data,
    % nodes near coords,
    % nodes near coords align={vertical},
    every node near coord/.append style={font=\scriptsize},	
    ]
    \addplot[plot 1] coordinates{(Spectra,0.591) (Surface,0.842)};
    \addplot[plot 2] coordinates{(Spectra,0.710) (Surface,0.884)};
    \addplot[plot 3] coordinates{(Spectra,0.725) (Surface,0.844)};
   \end{axis}
   
  \end{tikzpicture}
 \end{center}
 \captionsetup{justification=centering}
 \vspace{-3mm}
 \caption{Average performance per class of OSC+PRF+DR approach \\using Extra Tress model}
 \label{fig:OSC+PRF+DR_on_classes}
\end{figure}

The OSC+PRF+DR approach performed significantly better on Raman surface data compared to Raman spectra data across all classes on group 10-fold cross-validation, as shown in Figure \ref{fig:OSC+PRF+DR_on_classes}. For the Normal class, precision improved from 0.640 to 1.0, and the F1-score increased from 0.591 to 0.842. The Hyperglycemic class showed enhanced precision (0.674 to 0.792) and perfect recall (1.0), leading to an improved F1-score (0.710 to 0.884). Similarly, the Hypoglycemic class showed improved precision (0.712 to 0.826) and recall (0.774 to 0.864), resulting in a higher F1-score (0.725 to 0.844). 

\vspace{-3mm}
\vspace{-3mm}

Overall, the results indicate that Raman surface analysis delivers superior classification performance compared to Raman spectra analysis across all approaches. This further emphasizes the effectiveness of surface-based Raman data for improved classification performance in medical applications. In addition, OSC+PRF+DR consistently achieved the best performance, making it the most effective method in both scenarios.

%----------------------------------------
\section{Conclusion and Future Work}
\label{sec:conclusion}

In this study, we explored the use of machine learning for classifying exosomes derived from normal, hyperglycemic, and hypoglycemic cells based on their Raman spectra or surface. To improve classification accuracy, we applied several preprocessing and feature selection techniques, including OSC, OSC+PRF and OSC+PRF+DR. The results showed that OSC+PRF+DR, with Extra Trees as the classifier, consistently outperformed other methods, achieving the highest performance in both spectra and surface analyses. Specifically, OSC+PRF+DR resulted in an accuracy of 0.760, precision of 0.675, recall of 0.708, and an F1-score of 0.675 for spectra data, and an accuracy of 0.857, precision of 0.873, and an F1-score of 0.857 for surface data in a 10-fold cross-validation.

The study also found that Raman surface analysis provided superior classification results compared to Raman spectra analysis, highlighting its potential for medical applications. By integrating OSC, OSC+PRF, and OSC+PRF+DR, the approach improved spectral cleaning, noise reduction, and feature selection, thereby enhancing classification accuracy. These findings emphasize the promise of Raman spectroscopy and graph-based techniques for biomarker identification, suggesting that future work could refine feature selection and investigate outlier detection methods to further enhance performance.

Future work should focus on addressing potential overfitting to specific laboratory or spectrometer conditions, as all samples were sourced from the same machine. Acquiring additional samples under varied conditions would help assess model performance on independent data. Additionally, using nested cross-validation would further validate the OSC+PRF and OSC+PRF+DR, ensuring a more reliable evaluation of the models' generalization ability.

\section*{Acknowledgment}
\vspace{-3mm}
This study has emanated from research conducted with the financial support of Taighde Éireann – Research Ireland under Grant number 12/RC/2289\_P2.

%
% ---- Bibliography ----
\bibliographystyle{splncs04}
\bibliography{References.bib}

\end{document}